# Quantitative Evaluation of Microstrip, Dipole and L/C loop pTx Arrays for UHF MR Imaging

Debolina De[1], Aditya A Bhosale[1], Xiaoliang Zhang[1,2*]

Department of Biomedical Engineering, Department of Electrical Engineering, State University of New York at Buffalo, Buffalo, NY, United States

*Corresponding author: Xiaoliang Zhang, Ph.D.
Email: xzhang89@buffalo.edu

## Abstract

Ultra-high field MRI (7T+) unlocks a new era of brain research with superior resolution and signal-to-noise. Capturing intricate neural activity and detailed soft tissue pathology, this technology, coupled with advanced RF coil arrays, holds immense potential for clinical diagnosis and discovery. For high-field body imaging, microstrip resonators and dipole arrays show promise due to high-frequency operation and improved decoupling. Microstrip arrays excel in compactness, reduced radiation loss, and customizable lengths. Dipole arrays excel in deeper body penetration. Shorter wavelengths at ultrahigh fields can cause image inhomogeneity and elevated SAR. Multichannel transmit arrays address these issues. This research compares three common array types (L/C loop, microstrip, dipole) using CST Studio simulations to evaluate electric/magnetic fields and decoupling. This can guide RF array selection for ultrahigh field MRI.

## 1. Introduction

Magnetic resonance imaging (MRI) plays a powerful role in modern healthcare as a non-invasive diagnostic tool [1-39]. Compared to X-ray and nuclear medicine, MRI offers detailed soft tissue images without using harmful radiation. Traditionally, MRI scanners had a maximum field strength of 3T (whole body) or 7T (head/extremity) for clinical use. However, research is pushing boundaries with ultra-high field (UHF) scanners exceeding 7T, reaching 8T, 9.4T, and even 10.5T. [40-58]. This effort of pursuing higher magnetic field strength B0 is due to the intrinsic supra-linear increase in signal-to-noise ratio (SNR) in relation to the strength of the static magnetic field [59-61]. The SNR increase at UHF enables higher imaging resolution and unique imaging contrasts and enhanced opportunities for applications such as functional MRI (fMRI), spectroscopy, and susceptibility-weighted imaging (SWI) [62-73]. At the same time, there are challenges associated with UHF MRI: the increased radiofrequency (RF) power deposition in tissue, the required high operating frequency and consequently shortened wavelength, and the greater non-uniformity of the RF field [74-95]. By far the most popular approach to improve RF field homogeneity at UHF is a form of MRI called parallel transmission or 'pTx' [96-106].

## 1.1 RF Challenges at UHF

MRI uses radio waves to manipulate atomic nuclei and detect their response to create detailed images. A crucial component is the RF coil, which acts like a transmitter and receiver. It flips nuclei with radio pulses and picks up the resulting signal. These coils can be combined (transceiver) or separate (transmit & receive). For clear images, the transmit coil's radio waves need to be consistent throughout the scan area. The coils are tuned to a specific frequency based on the main magnetic field strength for optimal signal. As MRI field strengths increase, challenges arise due to shorter wavelengths and inhomogeneous fields. [107-113]. Imaging studies conducted at field strengths ranging from 4T to 9.4T showed that imaging could be done at these strengths and that basic problems like RF penetration and RF power needs wouldn't be very high at UHF [110, 114]. While measurements at frequencies as high as 220 MHz show that the power absorbed in conductive tissue rises with the square of the operating frequency [115], numerical estimates of the specific absorption rate (SAR) and absorbed power between 200 and 400 MHz indicate that this rise is not as rapid as low frequency approximations would have us believe [116]. Dielectric effects in the head are shown to outweigh eddy current shielding and RF penetration problems in simulations and tests [109, 117-120].

UHF MRI relies heavily on the electric field created by the RF coil, which heats the tissue and creates electric currents in conductive tissue in addition to the B1+ field inhomogeneity. Specific Absorption Rate (SAR), or the absorbed RF power per tissue mass, controls the amount of RF power deposited in tissue. The local electric field (E), local tissue conductivity ($\sigma$), and tissue mass density ($\rho$) are used to compute SAR.

$$SAR = \frac{\sigma ||E||^2}{2\rho}.$$

As a result, creating a strong and uniform B1+ field while controlling RF power deposition and adhering to FDA SAR regulations is the main engineering problem in the construction of an RF coil for ultra-high field MRI [121-136].

## 1.2 Parallel RF Transmission

Parallel RF Transmission was originally introduced to accelerate spin excitation speed, and now also used to mitigate transmit RF field inhomogeneity [96, 97, 99, 103, 104, 106, 123]. pTx produces distinct B1+ fields using a specialized RF coil with several channels. The spatially variable combined B1+ field made possible by these single-channel fields may be tailored for uniform imaging even at UHF with shorter RF wavelengths. In pTx, distinct B1+ fields are produced via a specialized RF coil featuring several separate channels. Even with the smaller RF wavelengths at UHF, these single-channel fields allow a spatially changing combined B1+ field that can be tuned for homogenous imaging [137, 138]. This is comparable to the widely recognized notion of parallel imaging at the receiving end, wherein the under sampled data is reconstructed through the utilization of several receive channels' sensitivities [97]. Due to the superposition of electromagnetic fields involved in pTx, SAR is significantly impacted [138]. The amplitude and phase of the currents to each array element can be freely adjusted using a parallel-transmit array coil, making it a crucial tool for mitigating B1+ inhomogeneity [77, 101, 123, 139, 140].

**1.3 Microstrip transmission line**

Microstrips offer distinct advantages over conventional coils in high-frequency RF coil designs for magnetic resonance imaging (MR) applications at ultrahigh fields [141-181]. These advantages include decreased radiation loss, a distributed element circuit, high-frequency capability, and less sample loading perturbation in the high-frequency range [81, 114, 159, 162, 182-197]. Composed of a narrow strip conductor, which can be either copper or silver, microstrips are purely dispersed. In microstrip, a low-loss dielectric material of a specific thickness divides the ground plane from the strip conductor. Microstrips are more straightforward to construct, have cheaper costs, no shielding requirements, and a higher Q-factor and superior decoupling performance [193, 198-201].

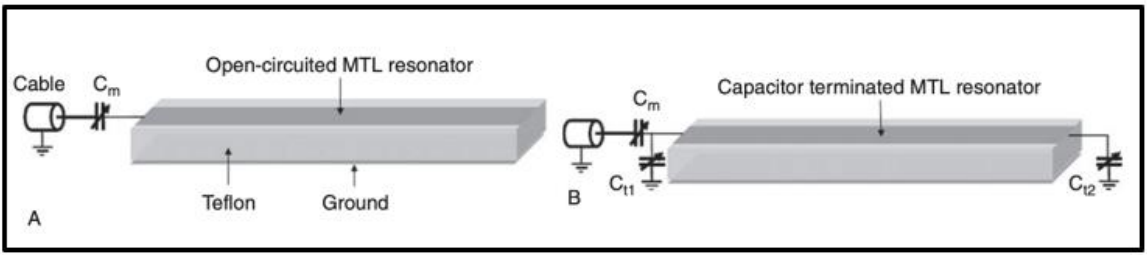

**Figure 1: Microstrip Transmission Line**

Figure 1 shows an RF resonator using a microstrip transmission line (MTL). Low-loss dielectric material, Teflon (Permittivity, εr =2.1 $F/m$ and conductivity, $\sigma = 0.6\ S/m$) which is the substrate of microstrip with height and length, and microstrip line composed of copper/silver. The microstrip line is used as λ/2 resonator with its ends terminated capacitors. The terminated shunt capacitors are used to reduce the physical length and to match the desired Larmor frequency (~300MHz, 7T) [193]. Like the ordinary MTL resonators, the capacitively terminated MTL resonators function similarly. The conductor's two ends or only one of them might be linked to the termination capacitors. The termination capacitors lengthen the electrical path to get the microstrip line to resonate at the intended frequency [193, 194, 198-202]. The resonant frequency of the MTL resonator terminated by one capacitor is calculated by using the following equation:

$$f_r = \frac{-1}{2\pi Z_0 (C_t)} tan(\frac{2\pi l \sqrt{\varepsilon_{eff}}}{c})$$

And the resonant frequency of the MTL resonator terminated on both sides is calculated using the following equation:

$$f_r = \frac{(2\pi f_r Z_0)^2 C_t C_{t_1} - 1}{2\pi Z_0 (C_t + C_{t_1})} tan(\frac{2\pi l \sqrt{\varepsilon_{eff}}}{c})$$

$Z_0$ is the characteristic impedance, $\varepsilon_{eff}$ is the effective permittivity, l is the length of the strip conductor, $C_t$ & $C_{t_1}$ are the termination capacitors [198].

**1.4 Half Wave dipole Antenna**

The most popular and extensively utilized type of antenna is the half-wave dipole. Radiating in a favorable pattern, it has a half-wavelength length. A conducting element, such as a wire or a metal tube fed in the center, makes up a half-wave dipole [203-223]. Variations in voltage and current are observed throughout the radiating area of the antenna. Because the half-wave dipole antenna's ends are often left open, the current at those points is always zero when the voltage is at its highest [200, 224]. The following equation can be used to calculate the resonant frequency of the half-wave dipole:

$$L = \frac{468}{f}$$

$$E = \frac{L}{2}$$

L is the total length of the dipole in feet, f is the frequency in MHz, and E represents the length of each dipole in feet.

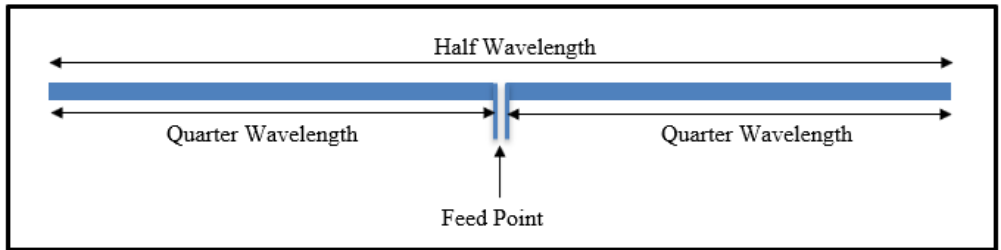

**Figure 2: Half wave dipole antenna**

**1.5 L/C loop coil**

Rectangular loop coils show promise for 7T MRI due to their compactness and efficient B1+ field generation. They can be arranged in arrays for parallel reception, potentially improving scan speed and signal quality. However, challenges include achieving uniform B1+ fields and complex decoupling in arrays, affecting image quality. Despite these limitations, they remain viable, especially for specific body parts. Ongoing research aims to improve their performance, but other coil designs like dipole or MTL arrays might be preferable depending on the application.[224].

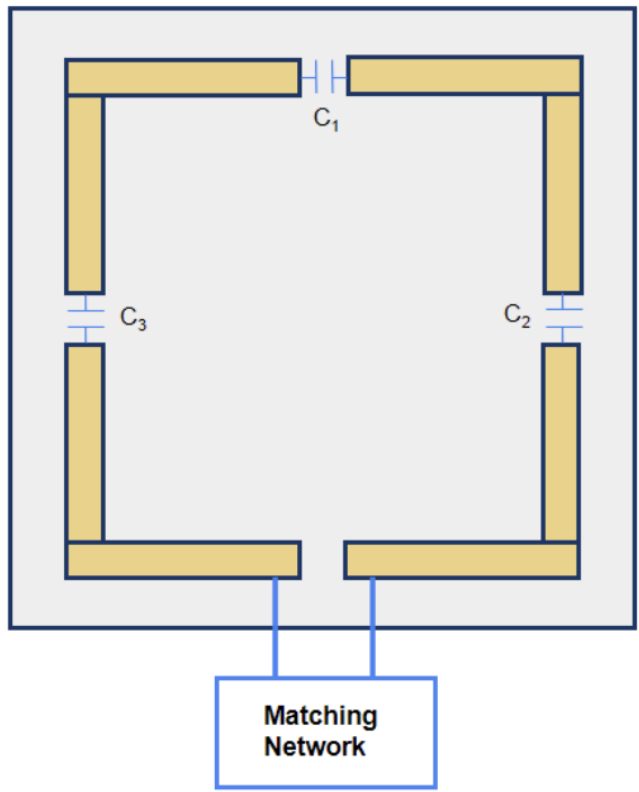

**Figure 3: Rectangular loop coil**

## 2. Methods

Using CST Studio Suite (Dassault Systèmes, Vélizy-Villacoublay, France), the numerical model of the 8-channel array was produced. Every element that is modeled has somewhat comparable external dimensions. We fixed the length of each antenna type to nearly 30cm for comparison. Other parameters, such as the meander length of the dipole, termination capacitors in the microstrip antenna, and capacitive elements in the rectangular surface coil, were modified to tune each of the three designs to 300MHz. Figures 4-6 show the designs. We have evaluated the magnetic field and electric field distributions of 3 different designs. Namely, two microstrip transmission arrays of different thicknesses of the dielectric material, a wave Meander dipole array, and an L/C loop surface coil at varying distances from the phantom. This distance is referred to as offset. For our research, we have used a cylindrical phantom with a radius of 20cm and length of 30cm, and it has the material properties of a human brain. We have also compared the decoupling performance of these arrays.

Input impedance matching, tuning, coil conductor, and decoupling circuits are the components of the coil circuits. The tuning, matching, and decoupling circuit components are described as ports to facilitate speedier simulations by leveraging the co-simulation capability. Lumped elements are the term used to describe the fixed circuit components. For the simulation environment to accurately capture coil losses, coil component losses such as the series resistance of the capacitors, inductors, and diodes must be included. More factors to consider are coarser mesh around the simulation domain's perimeter and local mesh improvements to guarantee that all ports and lumped parts are correctly connected to the coil conductors.

Figure 5 displays a snapshot of the 8-channel model of Microstrip Transmit line, figure 8 shows the 8-channel model of meandered dipole array and figure 11 shows the model of the 8-channel rectangular loop array. The fixed capacitors are shown as lumped elements (blue) while the matching and tuning circuits are represented as ports (red) in figure 7, 10, 12 and 13. This saves significant simulation time by allowing the coil parameters to be changed during circuit co-simulation without requiring a 3D re-simulation [225, 226]. The coils are loaded with a cylindrical phantom with known electrical properties ($\varepsilon = 51 F/m$, $\sigma = 0.4 S/m$). For building the coils the material of choice is copper, due to its excellent conductivity for RF currents. It offers good signal transmission and reception efficiency.

Ensuring a numerical model is resilient and trustworthy is the first thing to do after it is created. A finite integration solver is used with a tetrahedral mesh. Even though the simulation tools generate the mesh automatically, it is crucial to ensure the mesh has enough resolution for the model to prevent a staircase error [226, 227]. Adjusting the array S-parameters is the next stage after developing the numerical model of the transmit array. Appropriate circuit component values must be chosen to attain S11 values of less than −30dB for each array element. The overlap distance must also be optimized to reduce the coupling between neighboring components. The co-simulation domain allows for quick coil tuning and matching adjustments because these circuit components were modeled as ports. Until the minimal value of S21 is reached, the overlap optimization is done iteratively, requiring a fresh 3D simulation for each change in shape [226]. Between neighboring channels, there is a 45° incremental phase offset, and each coil element is driven with the same magnitude. Once the S-parameters are fully optimized, field maps can be generated from the fully tuned and matched system.

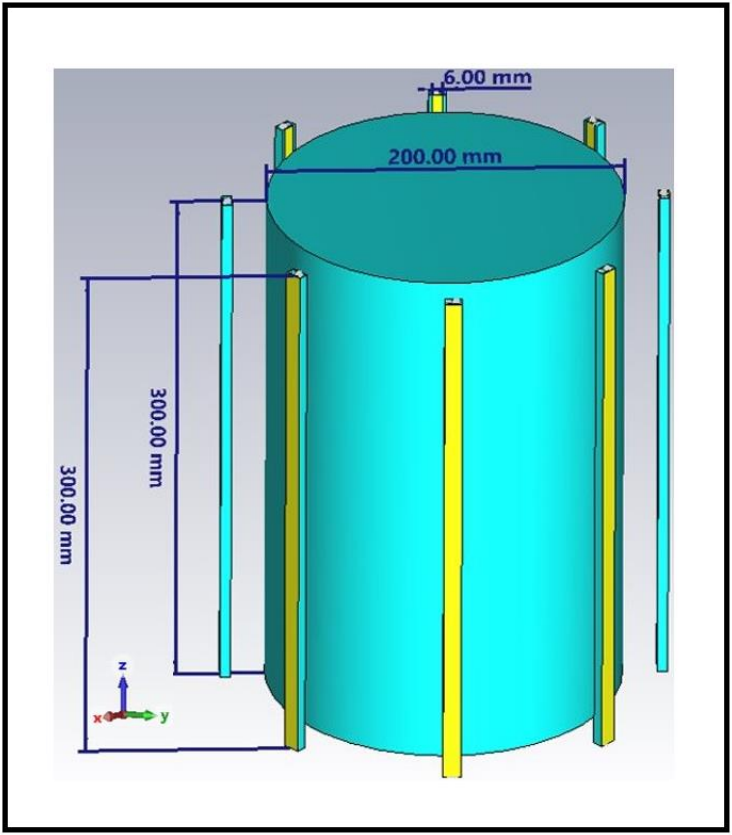

**Figure 4: Dimensions of 8-channel Microstrip Transmit Line Array loaded with cylindrical phantom with electrical properties of human brain tissue.**

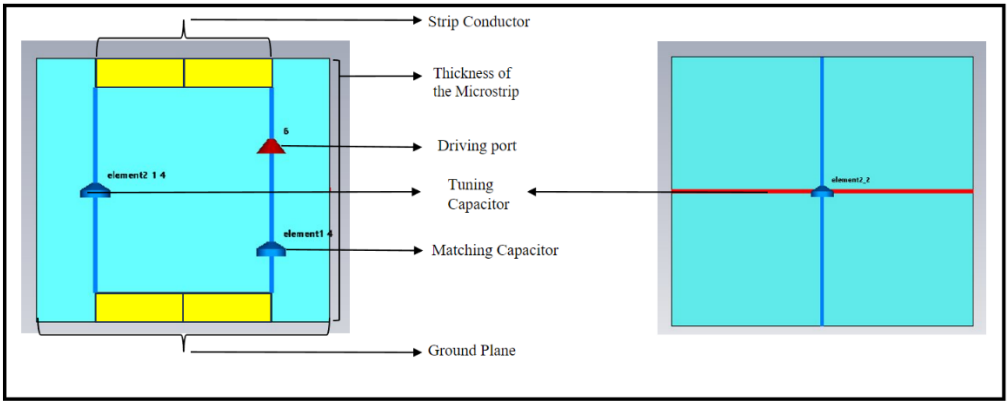

**Figure 5: (1) Microstrip top view with two lumped elements and the driving port (2) Microstrip bottom view with a single lumped element.**

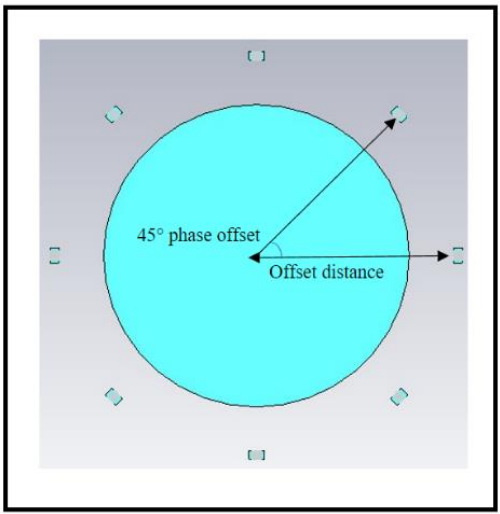

**Figure 6: Top view of 8-channel Microstrip Transmit Line Array loaded with cylindrical phantom with electrical properties of human brain.**

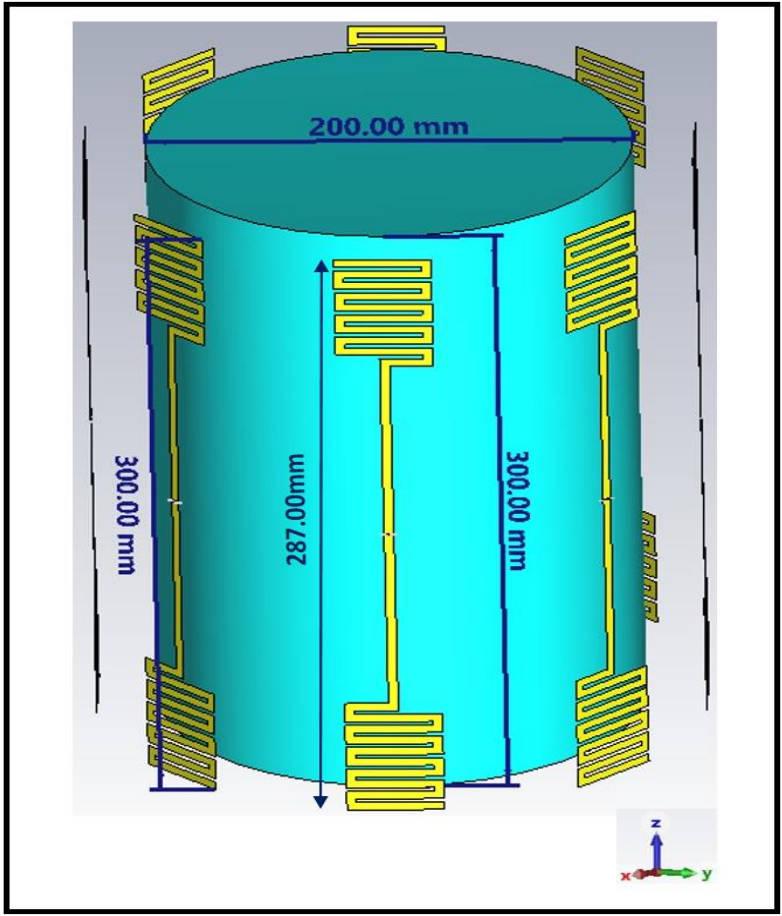

**Figure 7:** Dimensions of 8-channel Meander Dipole Array loaded with cylindrical phantom with electrical properties of human brain tissue.

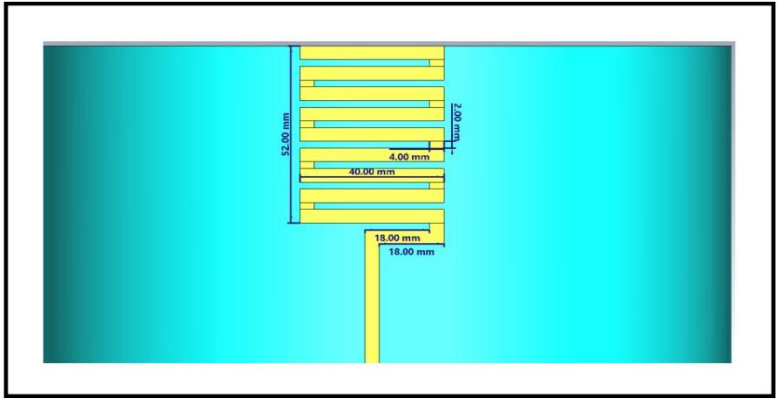

**Figure 8:** Dimensions of each meander of the dipole array.

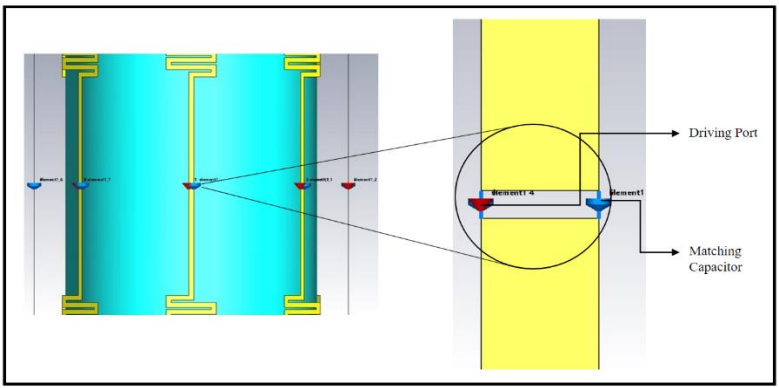

**Figure 9:** Feeding point and the matching capacitor are at the center of each dipole antenna.

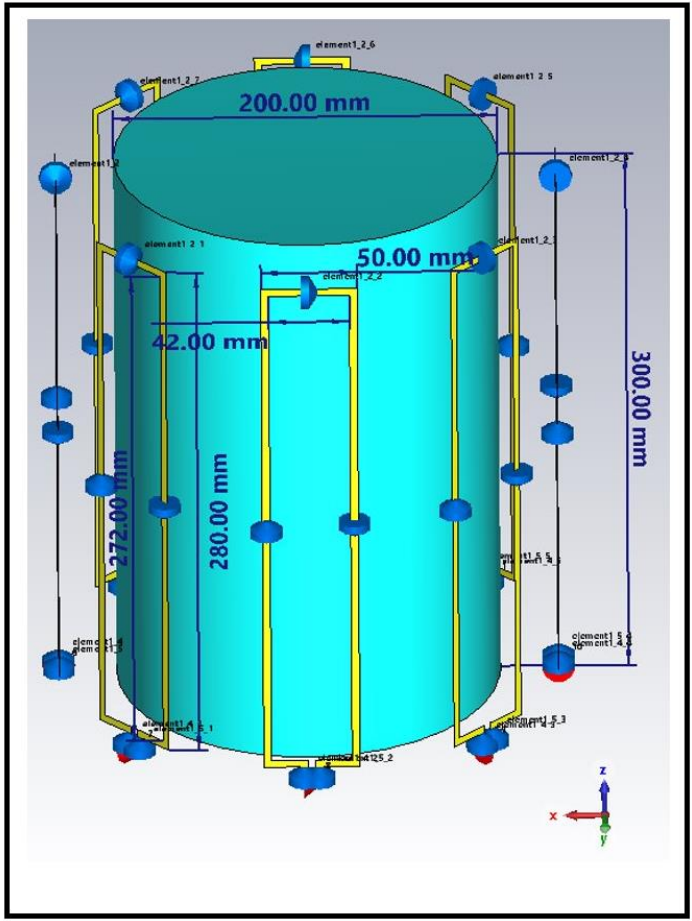

**Figure 10:** Dimensions of 8-channel rectangular loop array loaded with cylindrical phantom with electrical properties of human brain tissue.

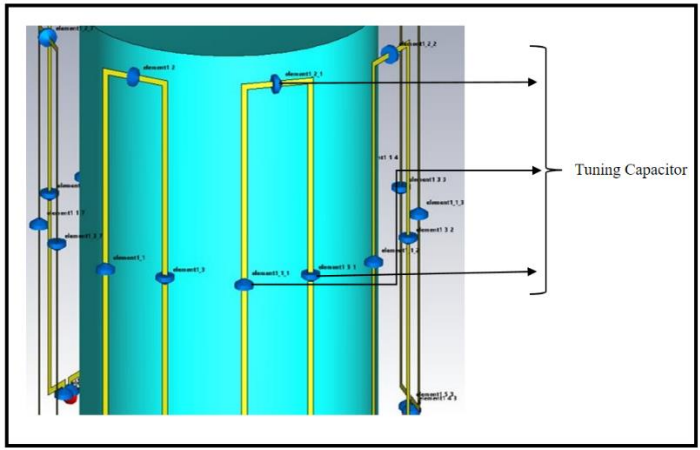

**Figure 5:** Tuning capacitor placement in a single element of a rectangular loop array.

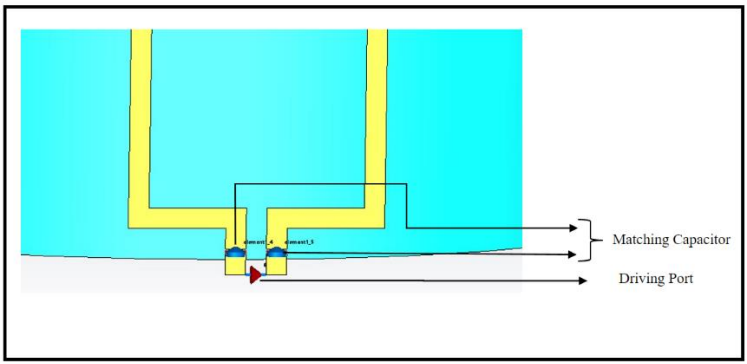

**Figure 12:** Matching circuit placement and the driving port for an individual rectangular loop.

# 3. Results

## 3.1 Magnetic Field Distribution

The B-field of a copper MTL antenna in MRI, induced by its E-field, penetrates the tissue phantom. Penetration depth depends on MRI frequency and tissue properties. This non-uniform B-field, crucial for nuclear excitation, is designed to focus on the ROI using a well-designed antenna. Teflon's low permittivity minimally affects the B-field compared to copper and tissue.

A copper dipole antenna in MRI generates a strong magnetic field in the center and weakens outwards. This field penetrates the tissue and is essential for exciting nuclei. A well-designed antenna concentrates this magnetic field within the region of interest for efficient signal generation and high-quality MRI scans, even though the inherent dipole design creates some unevenness.

A rectangular loop antenna concentrates the magnetic field within the loop due to current circulation and resonance. This magnetic field weakens rapidly outside and is strongest near the edges. While not perfectly uniform.

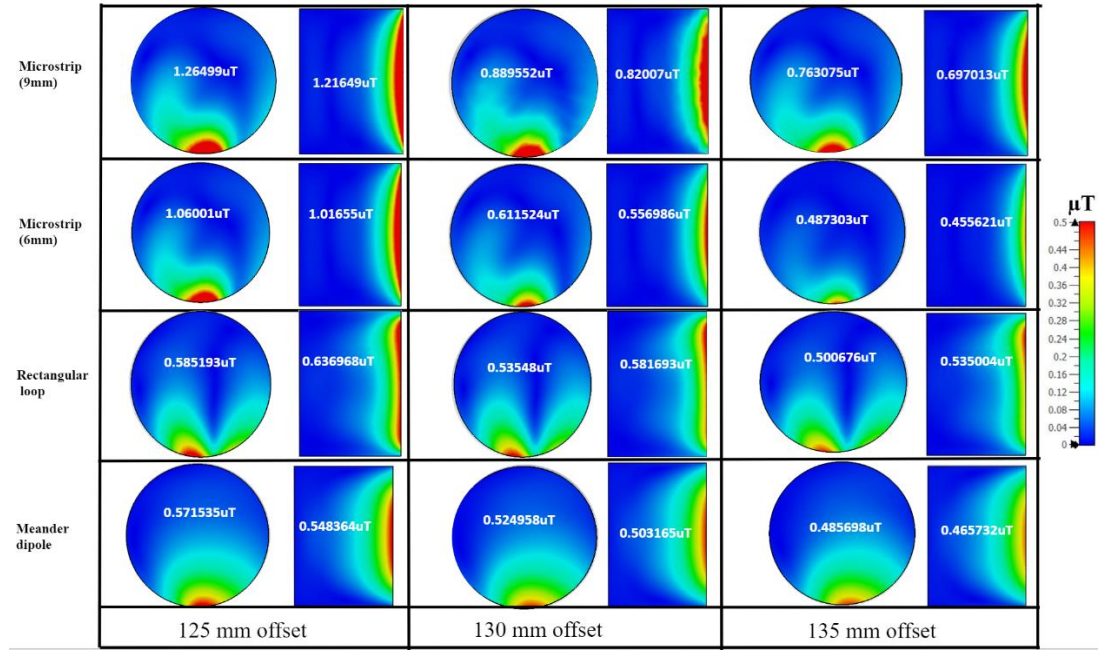

**Figure 13: Magnetic field distribution of a single channel / one element. (1) Microstrip with 9mm substrate thickness with varying offset distance. (2) Microstrip with 6mm substrate thickness with varying offset distance. (3) Rectangular loop with varying offset distance. (4) Meander Dipole with varying offset distance.**

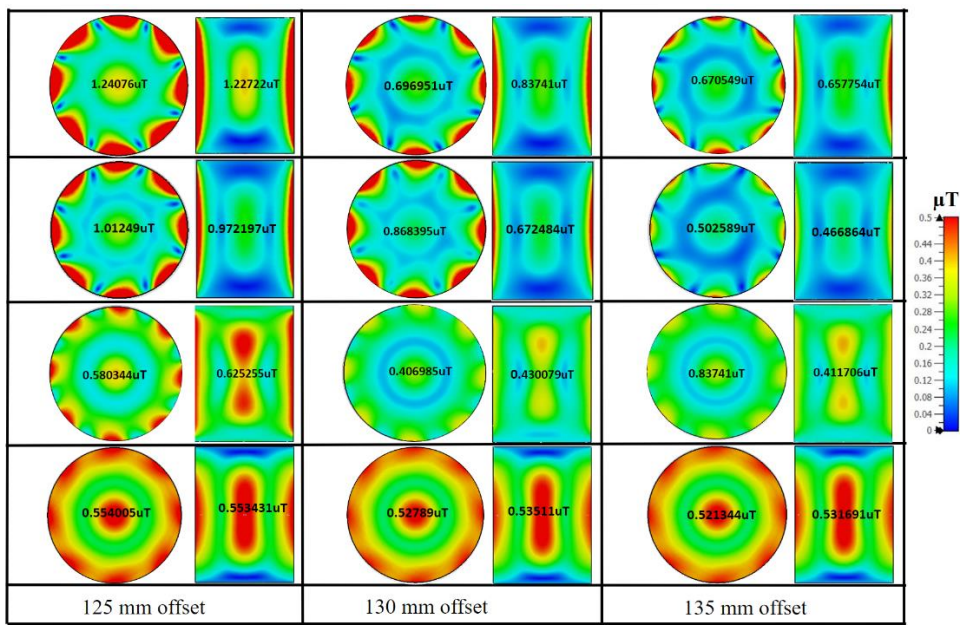

**Figure 14: Magnetic field distribution of 8 channel arrays. (1) 8 channel Microstrip with 9mm substrate thickness with varying offset distance. (2) 8 channel Microstrip with 6mm substrate thickness with varying offset distance. (3) 8 channel rectangular loop with varying offset distance.**

## 3.2 Electric Field Distribution

With some fringing extending outward, the electric field lines are mainly parallel to the microstrip conductor's length. When current passes through a conductor, the electric field changes direction. The electric field's strength is somewhat constant over the whole conductor surface. Due to the existing crowding effects, it tends to be slightly more robust on the margins.

A lozenge-shaped distribution of the highest electric field intensity surrounds the dipole's core. The electric field becomes weaker in every direction as we move away from the center. The electric field strength is meager at the dipole antenna's ends, where the two arms converge. It produces a radiation pattern appropriate for transferring energy into nearby tissue, predominantly in a plane perpendicular to the dipole axis.

The region that the rectangular loop encloses contains the highest concentration of electric field intensity. This is because resonance causes the circulating electric field that the current through the conductor amplifies. As we move outside of the loop's plane, the strength of the electric field swiftly decreases. This results from the electric field in the surrounding area being canceled out by the currents flowing in opposing directions along the rectangle. Unlike a dipole, the electric field inside the loop is not relatively uniform. Because there is a more significant concentration of current along the rectangle's edges, it tends to be stronger there.

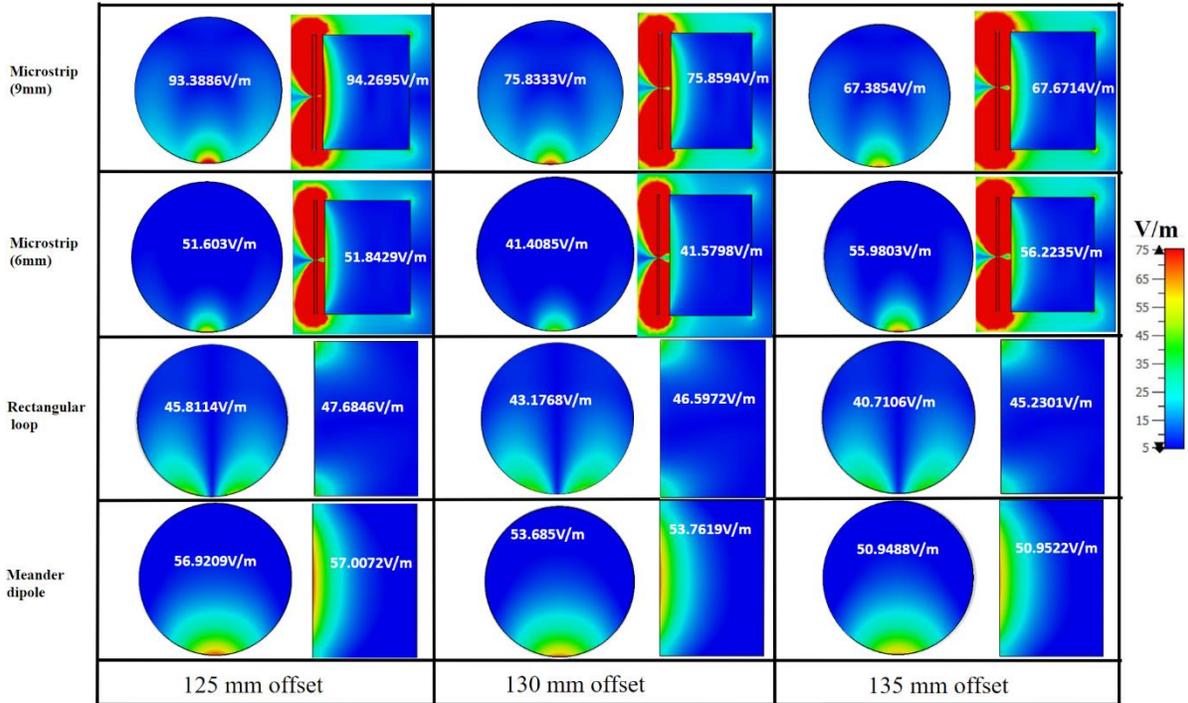

**Figure 16:** Electric field distribution of a single channel / one element. (1) Microstrip with 9mm substrate thickness with varying offset distance. (2) Microstrip with 6mm substrate thickness with varying offset distance. (3) Rectangular loop with varying offset. (4) Meander dipole with varying offset distance.

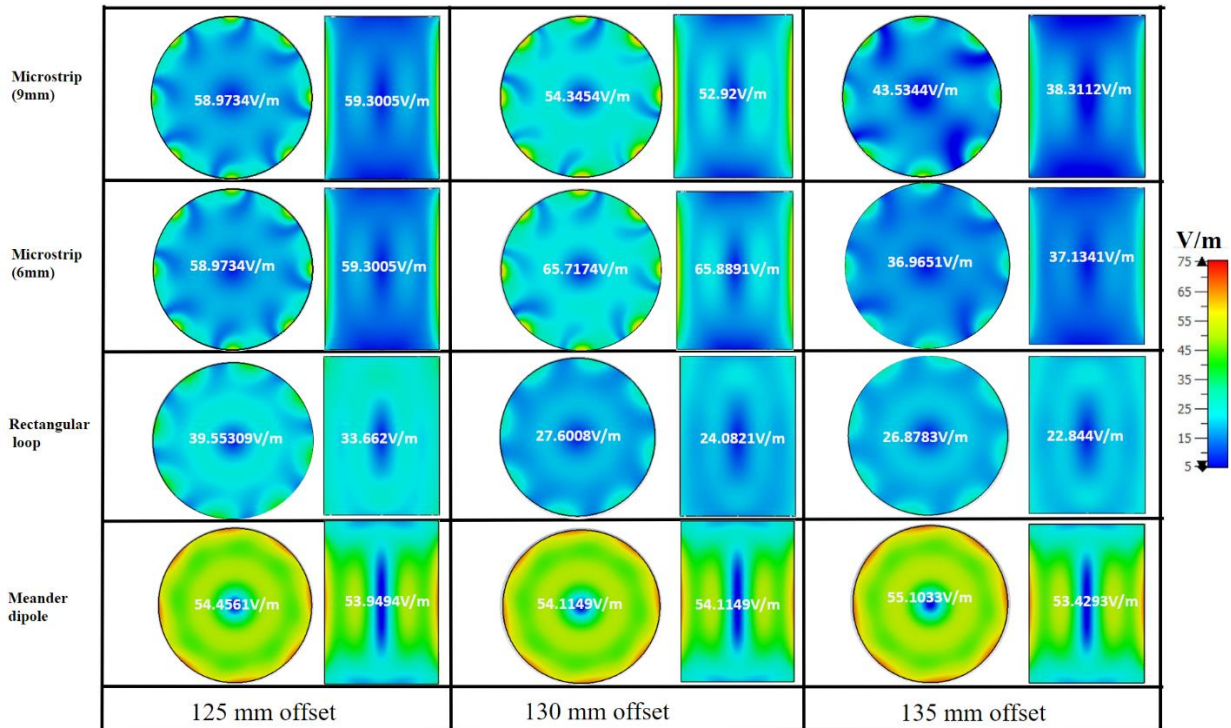

**Figure 17:** Electric field distribution of 8 channel array. (1) Microstrip with 9mm substrate thickness with varying offset distance. (2) Microstrip with 6mm substrate thickness with varying offset distance. (3) Rectangular loop with varying offset distance. (4) Meander dipole with varying offset distance.

## 3.3 B:E ratio

The B/E ratio provides a valuable indicator of how effectively an antenna couples energy into the surrounding medium in MRI. A higher B/E ratio generally signifies more efficient energy coupling and potentially deeper tissue penetration, while a lower B/E ratio suggests less efficient coupling and shallower penetration.

| Microstrip B:E | | Dipole B:E | Rectangular coil B:E |
|---|---|---|---|
| 6mm thickness | 9mm thickness | | |
| 0.021 | 0.014 | 0.01 | 0.013 |
| 0.015 | 0.012 | 0.009 | 0.012 |
| 0.009 | 0.011 | 0.008 | 0.012 |

Table 1: B/E ratios of Microstrip array with 6mm and 9mm substrate thickness, dipole array and rectangular loop array.

## 3.4 Q ratio (Unloaded Q-factor: Loaded Q-factor)

The Q ratio is a key metric for understanding the trade-off between bandwidth and power efficiency in an MRI antenna. A high Q ratio signifies a narrow bandwidth but potentially sharper B-field profile and lower power loss. A low Q ratio offers a wider bandwidth but might lead to increased power loss and a broader B-field profile [199].

| Offset | Microstrip Q-ratio ($Q_{unloaded}/Q_{loaded}$) | | Dipole Q-ratio ($Q_{unloaded}/Q_{loaded}$) | L/C loop Q-ratio ($Q_{unloaded}/Q_{loaded}$) |
|---|---|---|---|---|
| | 6mm | 9mm | | |
| 125mm | 1.87 | 3.14 | 1.005 | 1.3 |
| 130mm | 1.57 | 2.23 | 1.014 | 3.45 |
| 135mm | 1.42 | 1.82 | 1.015 | 2.46 |

Table 2: Q ratios of Microstrip array with 6mm and 9mm substrate thickness, dipole array and rectangular loop array.

## 3.5 Noise Correlation Matrix

The top row of figure 17 shows the noise correlation matrix of 8 channel Microstrip array and the bottom row shows 8 channel dipole arrays. For the microstrip array, crosstalk (S21) between the nearest neighbors was ranged from -18.7dB to -20db at varying offsets, whereas crosstalk (S21) between the next nearest neighbors had an average of -19dB. The crosstalk (S21) of the dipole array nearest neighbors' crosstalk (S21) between the nearest neighbors was ranged from -22dB to -28dB at varying offsets, and that of the next nearest neighbors had an average of -24.8dB.

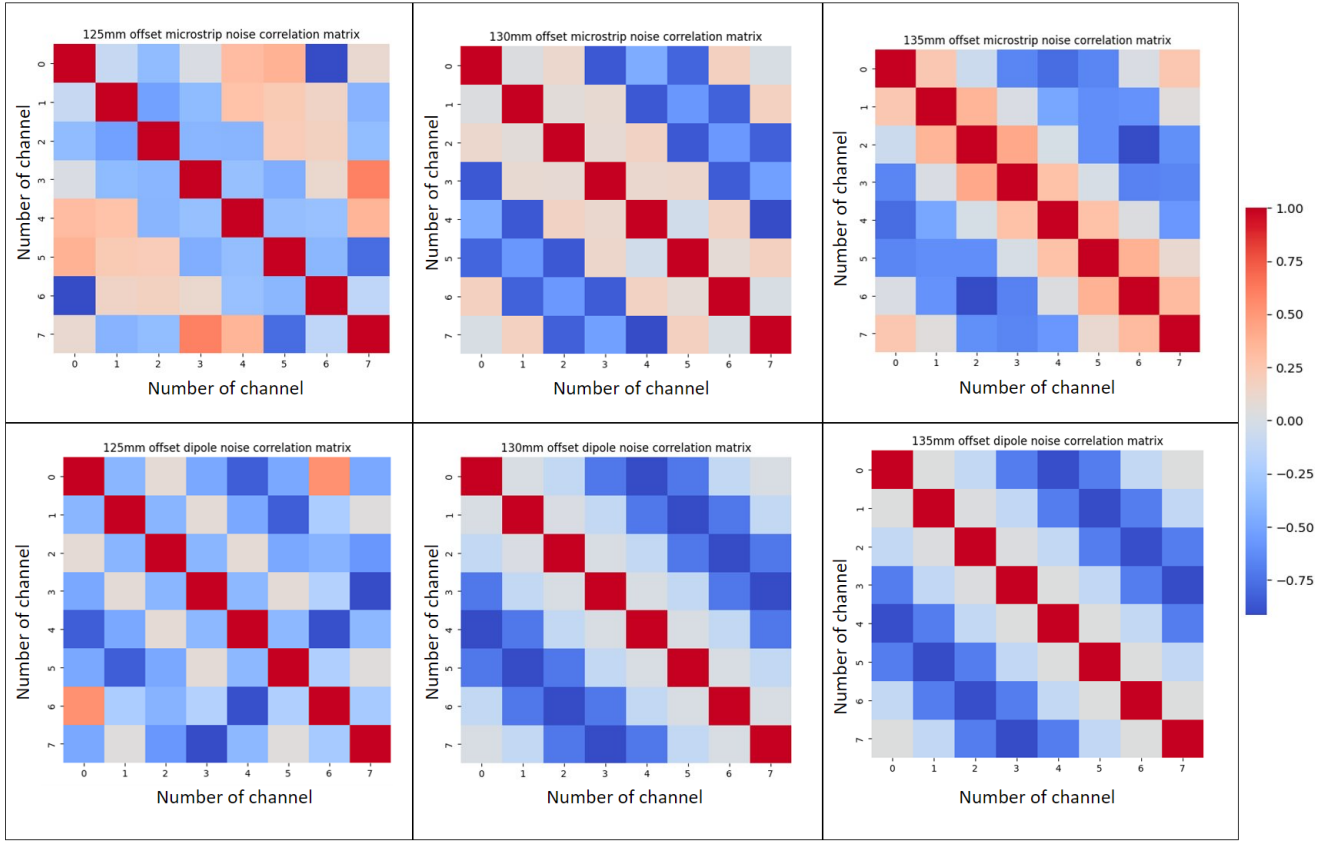

**Figure 17: Noise Correlation Matrix of 8 channel Microstrip (9mm substrate thickness) & 8 channel Dipole antenna with varying distance from the phantom (offset).**

# 4. Conclusion and discussion

In this study, we built and simulated 3 popular RF coil designs ultra-high field magnetic resonance imaging at 7 Tesla. 8 channel Microstrip transmit line array of 2 different substrate thickness, dipole array and rectangular loop array were designed and modeled. This included testing these 4 arrays placed at different distances from a phantom of electrical properties of a. As the distance increased from the phantom, the magnetic field penetration and strength decreased along with electric field distribution. From our simulations we see that microstrip has an in general higher magnetic field value. To optimize field strength within FDA safety limits, design modifications can be explored through simulations. Promising designs can then be validated by prototyping coils and testing them on a benchtop setup. The results can be used to further refine the simulation model, leading to a more iterative and efficient design process.

[145] B. Wu, X. Zhang, P. Qu, and G. X. Shen, "Design of an inductively decoupled microstrip array at 9.4T," *Journal of Magnetic Resonance,* vol. 182, no. 1, pp. 126-132, 2006, doi: 10.1016/j.jmr.2006.04.013.

[146] B. Wu, X. Zhang, P. Qu, and G. X. Shen, "Capacitively decoupled tunable loop microstrip (TLM) array at 7 T," *Magn Reson Imaging,* vol. 25, no. 3, pp. 418-24, Apr 2007. [Online]. Available: http://www.ncbi.nlm.nih.gov/entrez/query.fcgi?cmd=Retrieve&db=PubMed&dopt=Citation&list_uids=17371734

[147] Z. Xie, D. Xu, D. A. Kelley, D. Vigneron, and X. Zhang, "Dual-frequency volume microstrip coil with quadrature capability for 13C/1H MRI/MRS at 7T," *ISMRM Workshop on Advances in High Field MR, Pacific Grove, CA, USA,* 2007.

[148] Z. Xie and X. Zhang, "A novel decoupling technique for non-overlapped microstrip array coil at 7T MR imaging," in *the 16th Annual Meeting of ISMRM*, Toronto, Canada, P. o. t. t. A. M. o. ISMRM, Ed., May 3 -9 2008, p. 1068.

[149] Z. Xie and X. Zhang, "An 8-channel microstrip array coil for mouse parallel MR imaging at 7T by using magnetic wall decoupling technique," in *the 16th Annual Meeting of ISMRM*, Toronto, Canada, P. o. t. t. A. M. o. ISMRM, Ed., May 3 -9 2008, p. 2973.

[150] X. Yan, R. Xue, and X. Zhang, "Closely-spaced double-row microstrip RF arrays for parallel MR imaging at ultrahigh fields," *Appl Magn Reson,* vol. 46, no. 11, pp. 1239-1248, Nov 2015, doi: 10.1007/s00723-015-0712-1.

[151] X. Zhang, A. R. Burr, X. Zhu, G. Adriany, K. Ugurbil, and W. Chen, "A Dual-tuned Microstrip Volume Coil Array for Human Head parallel 1H/31P MRI/MRS at 7T," in *the 11th Scientific Meeting and Exhibition of ISMRM*, Miami, Florida, 2005, p. 896.

[152] X. Zhang, S. Ogawa, and W. Chen, "An inverted microstrip transmission line (iMTL) surface coil for 3T and 4T MR imaging," in *The 48th ENC*, Daytona Beach, FL, 2007.

[153] X. Zhang, R. Sainati, T. Vaughan, and W. Chen, "Analysis of single microstrip resonator with capacitive termination at very high fields," in *Proc Intl Soc Mag Reson Med*, UK, 2001, vol. 9, p. 699.

[154] X. Zhang, G. Shen, and B. Wu, "An optimized four-channel microstrip loop array at 7 T," in *ISMRM Meeting*, 2006, p. 2569.

[155] X. Zhang, K. Ugurbil, and W. Chen, "Microstrip RF surface coil design for extremely high-field MRI and spectroscopy," *Magn Reson Med,* vol. 46, no. 3, pp. 443-50., 2001.

[156] X. Zhang, K. Ugurbil, and W. Chen, "Microstrip RF Surface Coils for Human MRI Studies at 7 Tesla," in *Proceedings of the 9th Annual Meeting of ISMRM, Glasgow, Scotland*, 2001, p. 1104.

[157] X. Zhang, K. Ugurbil, and W. Chen, "A microstrip transmission line volume coil for human head MR imaging at 4T," *J Magn Reson,* vol. 161, no. 2, pp. 242-51, Apr 2003. [Online]. Available: http://www.ncbi.nlm.nih.gov/entrez/query.fcgi?cmd=Retrieve&db=PubMed&dopt=Citation&list_uids=12713976

[158] X. Zhang, K. Ugurbil, and W. Chen, "A volume RF coil using inverted microstrip transmission line (iMTL) for human head imaging at 7T," in *Proceedings of 11th annual meeting of ISMRM*, Toronto, Canada, 2003, vol. 11, p. 719.

[159] X. Zhang, K. Ugurbil, and W. Chen, "Method and apparatus for magnetic resonance imaging and spectroscopy using microstrip transmission line coils. 7023209," *US patent,* 2006.

[160] X. Zhang, K. Ugurbil, R. Sainati, and W. Chen, "An inverted-microstrip resonator for human head proton MR imaging at 7 tesla," *IEEE Trans Biomed Eng,* vol. 52, no. 3, pp. 495-504, Mar 2005. [Online]. Available: http://www.ncbi.nlm.nih.gov/entrez/query.fcgi?cmd=Retrieve&db=PubMed&dopt=Citation&list_uids=15759580

[161] X. Zhang *et al.*, "Human extremity imaging using microstrip resonators at 7T," in *Proceedings of the 21st Annual Meeting of ISMRM, Salt Lake City, USA*, 2013, p. 1675.

[162] X. Zhang, C. Wang, Z. Xie, and B. Wu, "Non-resonant microstrip (NORM) RF coils: an unconventional RF solution to MR imaging and spectroscopy," in *Proceedings of International Society for Magnetic Resonance in Medicine*, Toronto, Canada, 2008, vol. 16, p. 435.

[163] X. Zhang, Q. Yang, H. Lei, K. Ugurbil, and W. Chen, "Comparison Study of Microstrip Transmission Line (MTL) Volume coil and Shielded Birdcage Coil at 4T," in *Proceedings of the 10th Annual Meeting of ISMRM, Honolulu, Hawaii*, Honolulu, Hawaii, 2002, p. 881.

[164] X. Zhang, X. H. Zhu, and W. Chen, "Higher-order harmonic transmission-line RF coil design for MR applications," (in eng), *Magn Reson Med,* vol. 53, no. 5, pp. 1234-9, May 2005, doi: 10.1002/mrm.20462.
25